\documentclass[conference]{IEEEtran}
\IEEEoverridecommandlockouts

\usepackage{cite}
\usepackage{amsmath,amssymb,amsfonts}
\usepackage{algorithmic}
\usepackage{graphicx}
\usepackage{textcomp}
\usepackage{xcolor}
\usepackage{booktabs}
\usepackage{graphicx}
\usepackage{multirow}
\usepackage{threeparttable}    %
\usepackage{booktabs}
\usepackage{diagbox}
\usepackage{balance}

\def\BibTeX{{\rm B\kern-.05em{\sc i\kern-.025em b}\kern-.08em
    T\kern-.1667em\lower.7ex\hbox{E}\kern-.125emX}}

\begin{document}

\title{Perceptual Visual Quality Assessment: Principles, Methods, and Future Directions %\\
\author{\IEEEauthorblockN{Wei Zhou$^{1}$, Hadi Amirpour$^{2}$, Christian Timmerer$^{2}$, Guangtao Zhai$^{3}$, Patrick Le Callet$^{4}$, Alan C. Bovik$^{5}$ \\
$^{1}$Cardiff University, UK \quad
$^{2}$University of Klagenfurt, Austria \quad 
$^{3}$Shanghai Jiao Tong University,  China \\ $^{4}$University of Nantes, France \quad
$^{5}$University of Texas at Austin, USA
\\
Email: zhouw26@cardiff.ac.uk \quad hadi.amirpour@aau.at \quad christian.timmerer@aau.at \\ zhaiguangtao@sjtu.edu.cn \quad patrick.le-callet@univ-nantes.fr \quad bovik@ece.utexas.edu}}
}

\maketitle

\begin{abstract}
As multimedia services such as video streaming, video conferencing, virtual reality (VR), and online gaming continue to expand, ensuring high perceptual visual quality becomes a priority to maintain user satisfaction and competitiveness. However, multimedia content undergoes various distortions during acquisition, compression, transmission, and storage, resulting in the degradation of experienced quality. Thus, perceptual visual quality assessment (PVQA), which focuses on evaluating the quality of multimedia content based on human perception, is essential for optimizing user experiences in advanced communication systems. Several challenges are involved in the PVQA process, including diverse characteristics of multimedia content such as image, video, VR, point cloud, mesh, multimodality, etc., and complex distortion scenarios as well as viewing conditions. In this paper, we first present an overview of PVQA principles and methods. This includes both subjective methods, where users directly rate their experiences, and objective methods, where algorithms predict human perception based on measurable factors such as bitrate, frame rate, and compression levels. Based on the basics of PVQA, quality predictors for different multimedia data are then introduced. In addition to traditional images and videos, immersive multimedia and generative artificial intelligence (GenAI) content are also discussed. Finally, the paper concludes with a discussion on the future directions of PVQA research.
\end{abstract}


\section{Introduction}
The rapid development of multimedia communication has reshaped how we interact with digital content, with applications such as streaming services \cite{van2023tutorial}, video conferencing \cite{skowronek2014multimedia}, and immersive experiences in virtual reality (VR) \cite{balcerak2024immersive}. As these systems become more integral to modern communication, ensuring high-quality user experiences has become increasingly critical. Moreover, the perceptual visual quality of multimedia content directly influences user satisfaction. Therefore, it is important to develop effective perceptual visual quality assessment (PVQA) methods for various types of multimedia data, consisting of images, videos, VR content, and point clouds, among many others.

Because humans are the ultimate signal receivers of visual content, the most accurate way of accomplishing multimedia PVQA is to design and conduct subjective studies \cite{mantiuk2012comparison}. In this way, many subjective quality datasets have been proposed \cite{yue2023subjective,xu2021perceptual}, providing useful benchmarks for the development and comparison of computational models in PVQA.

Since subjective PVQA is time-consuming, expensive, and labor-intensive, it is of great interest to design computational models, known as objective quality models, that can automatically predict the perceptual visual quality of multimedia content. One of the earliest and simplest objective quality metrics is the peak signal-to-noise ratio (PSNR) or mean square error (MSE), which only relies on pixel-level differences. Although these metrics can be effective for certain tasks, such as evaluating codecs for the same content~\cite{huynh2008scope}, they often do not correlate well with human perception \cite{wang2009mean}.  The structural similarity index (SSIM) \cite{wang2004image} quantifies visual quality based on structural fidelity, while accounting for perceptual masking. Some variants, such as the multiscale SSIM (MS-SSIM) \cite{wang2003multiscale} and the feature similarity index (FSIM) \cite{zhang2011fsim}, have been proposed. While these conventional models assess signal fidelity, further improvement has been made via principles of both visual neuroscience and machine learning.

Other significant developments in objective PVQA are algorithms based on natural scene statistics (NSS) models \cite{mittal2012no,zhou2021no} and models of various aspects relating to the human visual system (HVS), such as visual attention \cite{liu2011visual}, contrast masking \cite{damera2000image}, and temporal vision \cite{moorthy2010efficient}. Since traditional models depend heavily on hand-crafted features, their adaptability and scalability may be limited. The emergence of deep neural networks (DNNs) has furthered the field by enabling the automatic extraction of perceptual features from human-labeled multimedia data \cite{zhou2021deep}. These models are particularly advantageous because they can learn complex patterns of human perception from large subjective quality datasets, improving their generalizability and robustness across diverse contents and distortions. More recently, the development of foundation models, such as large-scale transformers \cite{khan2022transformers} and multimodal large language models (MLLMs) \cite{wu2023multimodal}, has further progressed the model design of objective PVQA.
%by offering more powerful capabilities for quality prediction
%These models provide extraordinary flexibility, learning the complicated details of human perception in ways that traditional models could not.

At the same time, immersive multimedia and metaverse technologies are driving the rise of stereoscopic, light field, VR, point cloud, holography, and mesh content. Many new challenges arise in the development of objective PVQA models for such scenarios. Unlike traditional video or image content, immersive multimedia requires assessing multiple aspects of perceptual experiences, including depth perception, visual discomfort, and so on \cite{battisti2018toward}. These elements introduce new variables into the objective PVQA process, demanding innovative approaches beyond quality measurements on conventional image and video data. In addition, a large amount of generative artificial intelligence (GenAI) content has been recently produced by various generative models such as diffusion model \cite{ho2020denoising} and contractive language image pre-training (CLIP) \cite{radford2021learning}. However, there exist many quality issues related to GenAI content. For example, generated content may lose or poorly represent semantic information driven by input text prompts, posing challenges distinct from traditional PVQA methods used for natural content \cite{zhou2024adaptive}. Consequently, PVQA for immersive multimedia and GenAI content requires new methodologies that incorporate broader ranges of user experiences and contextual factors.

%generative adversarial network (GAN) \cite{goodfellow2014generative}

\begin{figure}[t]
	\centering
        \includegraphics[width=1.0\linewidth]{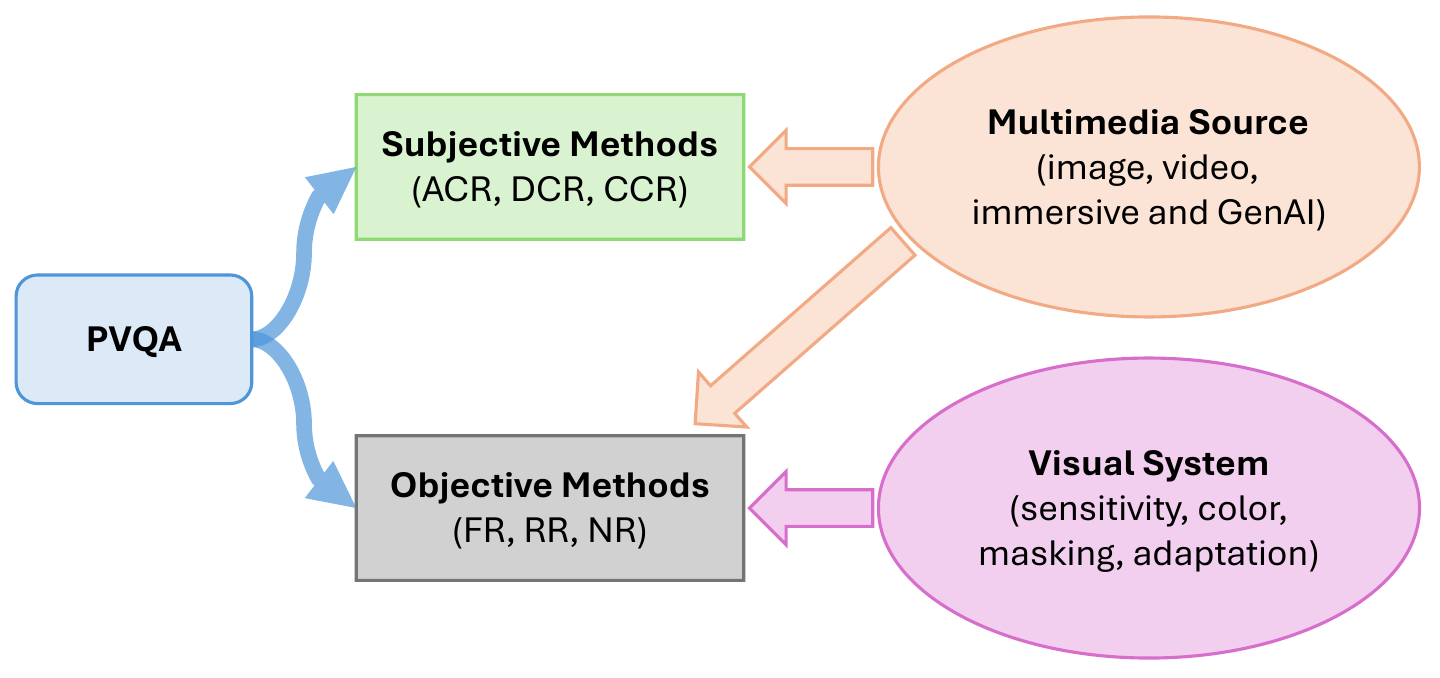}
        \vspace{-1em}
	\caption{The scope of perceptual visual quality assessment (PVQA).}
	\label{fig:fig1}
    \vspace{-1em}
\end{figure}

This paper aims to comprehensively review current trends and challenges in PVQA for multimedia communication. A depiction of the PVQA scope is shown in Fig. \ref{fig:fig1}. We begin in Section II by outlining the fundamental concepts of PVQA and its significance in multimedia contexts. Section III introduces existing PVQA methods for images and videos, comparing traditional objective quality metrics with more sophisticated perceptual models. Section IV addresses emerging challenges and highlights unique considerations of PVQA in immersive multimedia environments. In Section V, we explore the transformative role of foundation models in advancing the PVQA task. Finally, Section VI provides a conclusion and points out potential future directions for research in this rapidly evolving field.

\section{Perceptual Visual Quality Assessment}

PVQA is a critical research area that focuses on evaluating the perceptual visual quality of multimedia content based on human perception. Traditional metrics, such as the MSE, bitrate, or compression ratio, cannot accurately reflect user experiences because they do not account for the characteristics of the HVS. PVQA bridges this gap by incorporating insights from psychology, visual neuroscience, and computational modeling to assess content quality that aligns with human perception.

The proliferation of multimedia services, ranging from streaming platforms to VR and GenAI, has given rise to a significant need for tools that can help ensure perceptually satisfying user experiences while optimizing computation and bandwidth use. Reliable PVQA methods can improve the efficiency of both content delivery and user satisfaction.

\subsection{Visual Modeling} 
Modeling relevant aspects of the visual system plays a central role in the development of successful quality prediction models. Visual perception is highly adaptive, and understanding the way humans perceive varying resolutions, contrasts, and motion dynamics is important. To design appropriate subjective studies and build good PVQA prediction models, it is necessary to account for fundamentally key low-level properties of the visual system, including:

\textbf{Spatial and Temporal Sensitivity:} The visual system is sensitive to specific spatial frequencies, orientations, and motion dynamics. For example, it is sensitive to high-contrast edges and textures but less sensitive to smooth regions \cite{kelly1977visual}.

\textbf{Color Perception:} Human color perception is based on the interaction of cones in the retina, which roughly respond to red, green, and blue wavelengths \cite{wald1968molecular}. They are then efficiently encoded into color difference signals much like chroma signals in digital videos. While luminance changes often dominate visual quality perception, color information is also important.

\textbf{Masking Effects:} Many visual artifacts are less noticeable when embedded in high-energy patterns or high-contrast textures. This phenomenon, known as masking, is fundamental to the modeling of visual quality perception.

%used in compression techniques like JPEG 2000 \cite{zeng2002overview}. Moreover, in visual attention, masking can reduce the prominence of low-saliency areas in favor of high-saliency ones.

\textbf{Adaptation:} The HVS is able to adapt to varying viewing conditions, such as lighting, screen resolution, and viewing distance~\cite{amirpour2021impact}, which influence the perceived quality of images and videos.

\subsection{Subjective Quality Assessment} 

Subjective quality tests employ various methodologies to gather human feedback on the perceptual quality of visual content, each tailored to specific goals. The absolute category rating (ACR) method involves participants rating a single visual signal at a time on a predefined scale (e.g., 1 to 5). Although it can offer simplicity and efficiency, such a method is sensitive to context effects. The degradation category rating (DCR) method, a double-stimulus approach, asks viewers to assess the degradation level of an impaired content relative to a reference. This can increase sensitivity to distortion but does not correspond to usual real-world visual experiences. The comparison category rating (CCR) method, another double-stimulus approach, evaluates the relative quality of two contents (e.g., reference vs. impaired) by assigning scores to indicate preference or quality difference. Variants of these techniques are explored in detail in~\cite{pinson2003comparing}.
%The double-stimulus continuous quality scale (DSCQS) involves participants rating both reference and impaired content on continuous scales. The double-stimulus comparison scale (DSCS) directly compares visual pairs on a discrete scale, focusing on relative quality differences. In contrast, the single-stimulus continuous quality evaluation (SSCQE) allows dynamic quality ratings over time using a slider, providing high temporal resolution suitable for tracking rapid quality changes. SSCQE can be enhanced with hidden reference removal, ensuring ratings align with DSCQS outcomes. Each method has distinct strengths, such as ACR's simplicity, DCR's focus on impairment, DSCQS's robustness to context, and SSCQE's suitability for real-time evaluations. Combining these approaches offers insights into visual quality while balancing accuracy, efficiency, and viewer fatigue~\cite{pinson_comparing_2003}.

Subjective quality tests often produce metrics that capture and quantify user feedback, with two widely used examples being the mean opinion score (MOS) and the difference mean opinion score (DMOS). While the MOS provides a numerical summary of perceived quality by averaging participants' ratings, the DMOS focuses on the average difference between the scores assigned to high-quality reference stimuli and their corresponding distorted versions. By subtracting the reference stimuli's scores, DMOS reduces the influence of video content, offering a less biased measure of quality degradation.
To ensure reliability, confidence intervals (CI) are used to indicate the range within which the true MOS likely falls, accounting for variability among viewers. Another important concept is the just noticeable difference (JND)~\cite{amirpour2022between}, which measures the smallest change in visual quality that is perceptible to viewers, making it critical for studies focused on fine-grained quality adjustments. To improve the accuracy of gathered subjective data, outlier removal is often employed. Outliers, which may arise from participant errors or extreme subjective variability, can impair the value of subjective datasets. %Standardized methods such as analyzing z-scores or interquartile ranges help identify and exclude outliers, ensuring that the calculated MOS and CI reflect the consensus of participants more accurately. Together, MOS, CI, JND, and outlier management provide a robust framework for assessing and interpreting subjective test outcomes, ensuring both statistical rigor and perceptual relevance.

\subsection{Objective Quality Metrics} 

The evolution of objective quality metrics has progressed through distinct stages, driven by the need for cost-effective, accurate, and perceptually aligned evaluations to replace subjective tests. Generally, objective quality metrics are categorized into full-reference (FR), reduced-reference (RR), and no-reference (NR) approaches. FR metrics, such as PSNR and SSIM, compare impaired visual information to pristine references. RR metrics use partial reference information to balance accuracy and efficiency, while NR metrics assess quality without references, making them essential for real-world applications like streaming.

Initially, objective quality metrics rely on mathematical modeling to measure distortions but often lack perceptual relevance. To address this, models incorporating HVS features, like visual information fidelity (VIF) \cite{sheikh2005visual}, improve alignment with human perception. Machine learning techniques now play an important role, and data-driven approaches such as video multimethod assessment fusion (VMAF) are widely used. Modern advancements, including deep learning, self-supervised learning, and transformer architectures, have great potential and enable robust assessment across diverse contents and scenarios. The performances of all these techniques are typically evaluated in terms of correlation against the outcomes of subjective tests, using metrics such as Pearson correlation coefficient (PCC), rank-order correlation coefficient (RCC), coefficient of determination (R$^2$), mean absolute error (MAE), and root mean square error (RMSE).

\section{PVQA Methods for Image and Video}

\subsection{Principles}

Traditional objective PVQA is broadly divided into two categories: image quality assessment (IQA) and video quality assessment (VQA). IQA focuses on assessing the perceptual quality of still images, primarily relying on spatial information to detect distortions such as blurring, noise, and compression artifacts. In contrast, VQA addresses the assessment of video content, where both spatial and temporal dimensions play a critical role. Temporal information in VQA captures motion dynamics and frame-to-frame consistency, enabling the detection of distortions like judder, flicker, or temporal aliasing. While IQA serves as a foundational framework, the inclusion of temporal features in VQA introduces additional complexity, making it essential for applications involving dynamic visual information.

\subsection{Image Quality Assessment}
Traditional IQA approaches leverage spatial and transform perceptual principles, like those outlined above, to analyze the quality of visual content. Top-performing models such as SSIM, VIF, and VMAF incorporate key concepts such as bandpass statistical modeling and masking to accurately measure and predict perceived quality. 

In addition to spatial and transform domain techniques, learning-based IQA approaches have advanced significantly, leveraging recent breakthroughs in deep learning. For example, self-supervised learning methods, such as CLIP, as utilized in CLIP-IQA~\cite{wang2023exploring},  demonstrate remarkable potential. Similarly, transformer architectures, such as vision transformers, excel at capturing long-range dependencies and spatial relationships, enhancing their effectiveness in predicting image quality. Graph-based neural networks, such as GraphIQA \cite{sun2022graphiqa}, capture intricate dependencies between the regions of images. %Techniques like multi-task learning, used in IQA frameworks to predict both distortion type and overall quality \cite{}. 
Attention mechanisms, as in models like the Swin transformer, enhance feature focus on important regions, leading to better quality modeling. These advancements collectively enable highly accurate, scalable, and adaptable PVQA, addressing the growing complexity of modern multimedia systems.

%These techniques form the foundation for addressing common issues like compression artifacts, noise, and blurring, offering reliable quality assessments for general-purpose multimedia systems. 

% Traditional IQA approaches assess visual quality using spatial and transform domain techniques. Spatial methods focus on signal fidelity, structural similarity, and natural scene statistics (NSS) to predict perceived quality, while transform domain techniques analyze frequency components like wavelets and DCT to detect distortions.
% Learning-based IQA extends these methods with advancements like self-supervised learning (e.g., CLIP-Q), transformers (e.g., ViT), and graph-based neural networks to model complex dependencies. Multi-task learning enhances predictions by leveraging shared features, and attention mechanisms, as in Swin Transformers, to improve focus on critical regions. These innovations address issues like compression artifacts, noise, and blurring, ensuring accurate, scalable, and perceptually aligned quality assessments for modern multimedia systems.

\subsection{Video Quality Assessment}

VQA extends the principles of IQA by incorporating temporal information, making it essential for evaluating dynamic content. Unlike IQA, which focuses solely on spatial features, VQA accounts for the interplay between spatial and temporal dimensions to capture motion dynamics, frame consistency, and perceptual artifacts unique to video sequences. Temporal artifacts such as judder, flicker, stuttering, and frame drops are key challenges in VQA tasks, as these distortions significantly affect the perceived quality.
Traditional VQA methods often adapt IQA techniques, such as PSNR and SSIM, to video content by applying frame-wise assessments and aggregating results over time. However, these struggle to capture temporal coherence and motion-induced distortions. More advanced approaches combine spatial quality, motion features, and temporal factors into a unified framework, enhancing alignment with human perception.
Emerging VQA techniques utilize deep learning to tackle the complexities of video content, building on approaches similar to those used in IQA. 

%For example, transformer-based models, effectively model long-range temporal dependencies and motion dynamics. These architectures use attention mechanisms to capture subtle changes across frames, enabling robust evaluations of temporal artifacts. Similarly, self-supervised learning approaches for VQA, such as those leveraging contrastive learning, utilize large-scale video datasets to learn representations that generalize across diverse distortion types and content scenarios.

In streaming applications, VQA must consider factors like quality switching, latency, and rebuffering events, as these significantly influence overall quality. Sudden quality fluctuations or playback delays can diminish visual experiences, even if individual frames are of high quality, highlighting the need for metrics tailored to adaptive streaming dynamics. Hybrid models such as ITU-T P.1203 and P.1204~\cite{rao2020bitstream} integrate bitstream-level information, playback events, and perceptual quality predictions to provide comprehensive assessments. Additionally, simplified VQA models like VQM4HAS~\cite{amirpour2024real} leverage encoding information to efficiently predict VMAF scores for all representations in the bitrate ladder, addressing the computational expense of directly calculating VMAF for each representation.

%\subsection{Video Quality Assessment in Streaming and Compression}

\section{PVQA Methods for Immersive Multimedia}
The rise of immersive multimedia technologies, such as VR and 3D models, has transformed how users interact with digital content. Unlike traditional 2D multimedia, immersive experiences rely on more complex visual cues, making PVQA even more critical. To ensure enhanced user experience, the perceptual visual quality of such content should be assessed in ways that reflect both the medium’s unique properties and human perception. In this section, we discuss quality assessment methods specific to immersive multimedia, including stereoscopic content, light field, VR, and 3D models.

\subsection{Stereoscopic Content Quality Assessment}
Stereoscopic/3D (S3D) content is the generation and display of images or videos that create the illusion of depth. This is achieved by presenting two different images to each eye, mimicking the binocular disparity observed in real-world vision. Stereoscopic content presents unique challenges, such as vergence-accommodation conflicts and depth perception \cite{chen2017blind}. Additionally, more challenging asymmetric distortions may occur in left and right views. Therefore, stereoscopic visual quality assessment needs to consider both 2D-related factors and binocular mechanisms of stereoscopic vision. 

Compared with the traditional 2D case, subjective methods for stereoscopic content typically may involve asking users to rate their experiences along multiple dimensions such as image/video quality, depth perception, and overall satisfaction \cite{zhou20163d}. Early S3D FR QA models directly derive from existing 2D quality models. For example, by applying 2D algorithms to left and right views separately, and then integrating the outcomes with depth predictions of stereoscopic visual quality \cite{benoit2009quality}. More advanced algorithms have been proposed that model properties of binocular vision \cite{jiang2020full}.

However, since the original content is not always available, NR S3D  quality assessment methods for stereoscopic content have also been devised. These models employ the discriminative features of distorted stereoscopic content to evaluate perceptual S3D quality based on perceptual models such as 3D NSS as well as deep learning models \cite{zhou2019dual,xu2020binocular}.

\subsection{Light Field Quality Assessment}
A light field is a representation of the amount of light traveling in every direction through every point in space, capturing more information than traditional 2D images or videos. Light fields can provide several depth cues, including monocular cues, binocular cues, motion parallax, and refocusing. This presents challenges related to the density of viewpoint samples, view synthesis quality, and focus smoothness. Since light fields require a significantly higher amount of data as compared to traditional 2D content, effective compression while maintaining quality is a critical issue~\cite{amirpour2022advanced}.

Moreover, measuring the subjective quality assessment of light field content requires specialized viewing environments \cite{shi2018perceptual,amirpour2019reliability}, where users can freely explore the scene from multiple views. Common measurements used in this domain include depth perception accuracy, focus transition smoothness, and overall visual appeal. Objective methods for light field content have been devised based on spatial-angular measurement \cite{shi2019no}, tensor theory \cite{zhou2020tensor}, micro-lens images \cite{luo2019no}, etc.

\subsection{VR Quality Assessment}
VR applications have made great progress, providing new ways for immersive consumers to visualize and interact with visual information. Unlike conventional multimedia formats, VR systems employ head-mounted displays or CAVE automatic virtual environments. In this direction, 360-degree omnidirectional content has attracted significant research. Many specific distortions may be introduced throughout the processing chain of 360-degree omnidirectional content, leading to visual quality degradations.

In the literature, there exist a few subjective datasets for omnidirectional content \cite{duan2018perceptual}, and they suffer from some critical issues, including restricted sizes, insufficient content, diversity, and limited distortion representation. Meanwhile, existing objective QA prediction models either project 360-degree content onto 2D planes, and then apply traditional 2D quality assessment methods \cite{zhou2023blind}, or exploit convolutional neural networks to train deep models \cite{xu2020blind}, with little consideration of key aspects of the perception relating to 360-degree omnidirectional content, such as resolution, frame rate, motion sickness, visual attention, and the interactions among multiple viewports. Due to fluctuating network conditions, latency issues, and irregular head movements, effective adaptive bitrate algorithms to achieve optimal user experiences in 360-degree video streaming remain challenging.

% for which accurate PVQA models and effective perceptual optimization and decision-making methodologies are lacking

\subsection{Quality Assessment of 3D Models}
The evolution of 3D technologies has led to the widespread use of point clouds, meshes, digital heads (avatars), and emerging techniques like 3D Gaussian splatting. As these technologies become increasingly integrated into multimedia communication, ensuring the perceptual quality of 3D data becomes critical for maintaining user satisfaction.

Point clouds, meshes, digital heads, and 3D Gaussian splatting each represent different ways of structuring and rendering 3D objects, with each having distinct advantages and challenges when it comes to PVQA. Point clouds consist of a large number of discrete data points in 3D space. While they provide highly detailed and accurate representations of physical objects, their quality is often influenced by issues such as low point density and noise. Meshes, which connect the data points from point clouds with edges and faces, offer complete and structured models, but they can suffer from possible poor triangle distribution and texture mapping issues. Digital heads, or avatars, require high levels of detail and realism, especially of facial features and animation. The quality of these models is influenced by both geometric accuracy and texture fidelity, with subjective perception heavily depending on naturalistic facial movements and expressions.

Recently, 3D Gaussian splatting has emerged as an innovative technique for rendering complex 3D shapes, especially in real-time graphics and interactive applications. Different from traditional mesh-based rendering, 3D Gaussian splatting represents 3D objects as distributions of 3D Gaussians that provide advantages in terms of efficiency and flexibility in handling complex geometries. However, similar to other 3D representations, 3D Gaussian splatting also has quality concerns, including rendering artifacts and lack of detail.

The evaluation of these 3D representations requires robust PVQA methods, which can be broadly categorized into subjective and objective approaches. Subjective QA \cite{liu2022perceptual} often involves human participants rating the quality of 3D models based on visual realism, detail, smoothness, and other perceptual factors. These assessments are vital for understanding user preferences, particularly in immersive and interactive applications. Objective quality metrics \cite{zhou2024blind,sarvestani2024perceptual} employ computational methods to quantify quality attributes such as geometric accuracy, texture mapping, surface smoothness, and rendering efficiency. These metrics can provide reproducible and consistent results that are essential for automating the evaluation of subjective quality datasets and supporting real-time applications.

\section{PVQA in The Foundation Model Era}
The foundation model era marks a significant shift in the way advanced AI systems are designed and applied to multimedia communication. The term “foundation model” refers to large pre-trained models exposed to very large amounts of data, such as BERT \cite{kenton2019bert} and DALL-E \cite{ramesh2021zero}, that serve as the underlying framework for a wide variety of downstream applications. These models, which are often based on transformer architectures, are trained on vast datasets and can be fine-tuned for specific tasks, including quality assessment in multimedia communication.

Foundation models are increasingly capable of understanding complex relationships between multimedia content and human perception. They leverage extensive training on multimodal data, such as images, videos, texts, etc., to learn high-level representations that go beyond traditional feature extraction \cite{lu2019vilbert}. This paradigm shift has profound implications for quality assessment, as it offers more robust and context-aware tools for evaluating multimedia quality from a human-centric perspective.

Apart from utilizing MLLMs to better predict the perceptual visual quality of diverse multimedia data, GenAI content, such as deepfakes, requires specialized quality metrics. The quality of GenAI content is fundamentally different from that of conventional multimedia content, and thus, its evaluation requires new approaches that exceed pixel-based analysis \cite{zhang2024quality}. Therefore, numerous new quality assessment benchmarks and methodologies have been developed to evaluate the multimedia data generated by these large models. Subjective quality datasets have been built for GenAI content, which consist of text-to-image generation \cite{li2023agiqa}, image-to-image generation \cite{yuan2023pku}, and text-to-video generation \cite{kou2024subjective}. In these quality datasets, unique aspects of GenAI content, such as authenticity and consistency, are taken into account.

Early objective quality assessment models targeting GenAI content involve non-perceptual measures, like the inception score \cite{salimans2016improved} and Fréchet inception distance \cite{heusel2017gans}, hence are not able to predict the actual user preferences. Recognizing the unique challenges posed by the GenAI content production processes, several quality assessment models have been specifically developed for GenAI content, e.g., using MLMMs and CLIP-based methods \cite{fu2024vision}.

\section{Conclusion and Future Directions}
We have provided a high-level overview of PVQA in multimedia communication. This paper has broadly outlined both subjective and objective methods for evaluating perceptual visual quality. While subjective methods remain the gold standard for assessing user experience, objective quality models based on measurable factors are gaining significant attention, especially in large-scale and real-time applications where human evaluation is impractical. Furthermore, we have explored how quality assessment methodologies have evolved to account for emerging multimedia formats, including immersive multimedia content and GenAI images as well as videos, both of which present unique challenges due to their complexity and interactivity. The introduction of AI and foundation models into PVQA is a significant development, offering the potential for more accurate and automated models. These technologies have the potential to bridge the gap between subjective experiences and objective metrics, especially in highly dynamic or immersive environments. However, challenges remain in developing universally applicable models that account for a broad range of content types, viewing conditions, and distortions.

In the future, PVQA will likely focus on refining objective models for new forms of multimedia, improving the alignment between objective models and human judgments, and addressing the impact of increasingly complex content like augmented reality. Moreover, as GenAI models become more prevalent, the need will arise for specific PVQA models that can evaluate not only the technical quality of generated content but also perceptual realism and consistency. Future work will need to focus on developing more efficient models that reduce computational complexity while maintaining accuracy, enabling real-time evaluation across diverse multimedia contents and platforms.

% new multimedia; new protocols to asses new display, etc., Practical deployment; computational complexity for some use cases, [for benchmarking it is ok] for example, in production in streaming; dedicated hardware accelerator;

\balance
\begingroup
\renewcommand{\baselinestretch}{0.86}
\bibliographystyle{IEEEtran}
\bibliography{sample-base}%reference
\endgroup

\end{document}